\documentclass[submission,copyright,sharealike]{eptcs}
\providecommand{\event}{ThEdu'21}

\usepackage[T1]{fontenc}
\usepackage[utf8]{inputenc}
\DeclareUnicodeCharacter{92}{'}
\usepackage{amsmath}
\usepackage{amsthm}
\usepackage{listings}
\usepackage{multirow}
\usepackage{array}
\usepackage[export]{adjustbox}
\usepackage{breakurl}
\usepackage{underscore}
\usepackage{subcaption}
\usepackage{tabularx}

\newcommand{\smtlib}[0]{SMTLib}

\newcommand{\set}[1]{\{#1\}}
\newcommand{\lookup}[2]{lookup(#1, #2)}
\newcommand{\update}[2]{update(#1, #2)}
\newcommand{\enter}[1]{enter(#1)}
\newcommand{\leave}[1]{leave(#1)}
\newcommand{\addconstraint}[2]{addconstr(#1,#2)}
\newcommand{\unwindE}[2]{T_E(#1, #2)}
\newcommand{\unwindS}[2]{T_S(#1, #2)}
\newcommand{\return}[0]{\_\_ret}
\newcommand{\ASSERT}[0]{\text{\texttt{ASSERT}}}
\newcommand{\domain}[0]{\mathcal{D}}
\newcommand{\lequiv}[0]{\,\approx\,}
\newcommand{\lnequiv}[0]{\,\not\approx\,}
\newcommand{\impl}[0]{\,\Rightarrow\,}
\newcommand{\ithen}[2]{\text{\texttt{if} }(#1)\,#2}
\newcommand{\ite}[3]{\text{\texttt{if} }(#1)\,#2\,\text{\texttt{else}}\,#3}

\newcommand{\while}[2]{\text{\texttt{while} }(#1)\,#2}
\newcommand{\tloopr}[2]{\text{\texttt{LOOP}}(range(#1,#2));}
\newcommand{\returnstmt}[1]{\text{\texttt{return} }#1}
\newcommand{\tuple}[1]{\langle{#1}\rangle}

\newcommand{\tatsu}[0]{{\sc Tatsu}}
\theoremstyle{definition}
\newtheorem{definition}{Definition}

\title{Automated Instantiation of Control Flow Tracing Exercises\footnote{~We would like to thank Stefan Podlipnig for his continuing support and input on the paper. His lead in the introductory programming course for computer science on TU Wien with his ideas for \tatsu{} made this work possible. We would also like to thank Nikolaj Bj\o{}rner for his help with integrating Z3 in \tatsu{}.}}
\author{
Clemens Eisenhofer
\institute{TU Wien\\Vienna, Austria}
\email{clemens.eisenhofer@tuwien.ac.at}
\and
Martin Riener
\institute{TU Wien\\Vienna, Austria}
\email{martin@derivation.org}\\
}

\def\titlerunning{Automated Instantiation}
\def\authorrunning{C. Eisenhofer, M. Riener}

\begin{document}
\maketitle
\begin{abstract}
  One of the first steps in learning how to program is reading and tracing existing code. In order to avoid the error-prone task
  of generating variations of a tracing exercise, our tool \tatsu{} generates instances of a given code skeleton automatically.
  This is achieved by a finite unwinding of the program in the style of bounded model checking and using the SMT solver Z3 to find models for
  this unwinded program.
\end{abstract}

\section{Introduction}
Amongst the aims of introductory programming courses is to teach the understanding of small programs. Lister et al.~\cite{DBLP:conf/ppig/TeagueL14,10.5555/2459936.2459938}
argue that students are often able to trace a program before they can formulate its purpose. This indicates that tracing exercises help students to
reach an operational phase where they can reason about their programs. Lister also proposes multiple choice questions~\cite{DBLP:conf/sigcse/Lister01} that ask the students to identify the output of a small program correctly. 
Other exercises require the students to identify which code snippets should fill a hole in a program skeleton such that it follows a given specification or ask whether some proposed modifications of a given program lead to an error.

Even when an introductory programming course does not fully subscribe to this ``trace before writing your own code'' view, it can complement other teaching methods. The authors are involved in such a course for computer scientists on TU Wien with about 650 students. There, tracing questions are used for practice, but also contribute 20\% to the overall grade in form of an online assessment within TUWEL, the local Moodle instance. The larger part of the course is spent in lab sessions explaining self-written programs to the instructor and the group.

Since the high number of students requires an adequate pool of questions we create variations of a single question by instantiating a code schema with different values. Historically this has been done manually which is error prone. As an alternative these values can be generated with an SMT solver using ideas borrowed from bounded model checking. For \tatsu{} we use the Z3 solver~\cite{z3} as it supports some non-standardised functionalities required by our tool. In the following we report on the ongoing development of \tatsu{} and its use to generate quizzes. These quizzes contain single/multiple choice questions as well as questions asking for the output or return value of a given function. The program generator is open source and can be downloaded from our GitLab instance\footnote{~\url{https://git.logic.at/ep1-tools/tatsu-generator}}.

Generating programs for such exercises in this context is quite different from general synthesis or verification tasks: the programs are short and beginners are not confronted with the full syntax of the language. We also focus on questions of the kind ``How often is this loop executed?'' rather than checking for corner cases like integer overflows. \tatsu{} automatises the task by taking the skeleton of a Java program together with a set of pre- and post-conditions and generating a set of instances that fulfill these conditions.

\section{Overview}
\begin{figure}[ht]
  \centering
  \begin{minipage}[t]{0.40\linewidth}
    \includegraphics[width=\textwidth,valign=t]{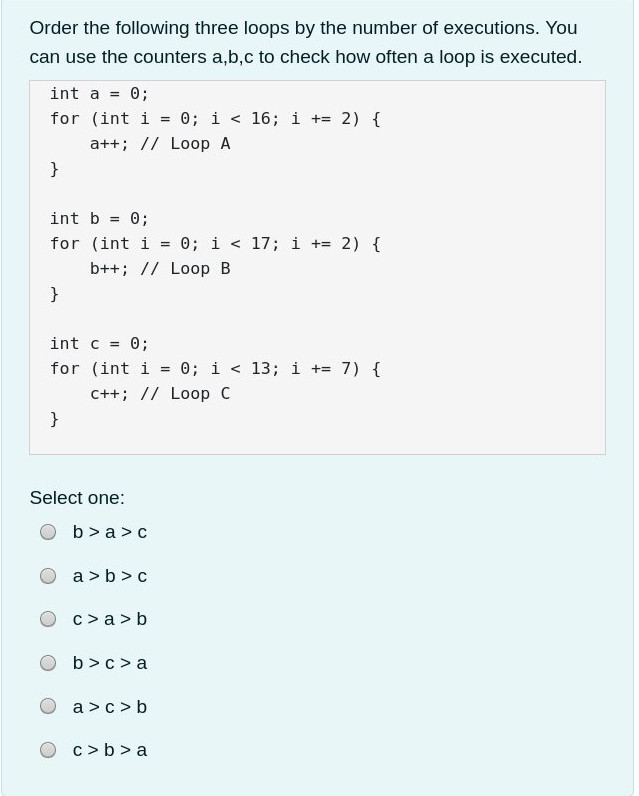}
  \end{minipage}
  \hspace{4mm}
  \begin{minipage}[t]{0.40\linewidth}
    \includegraphics[width=\textwidth,valign=t]{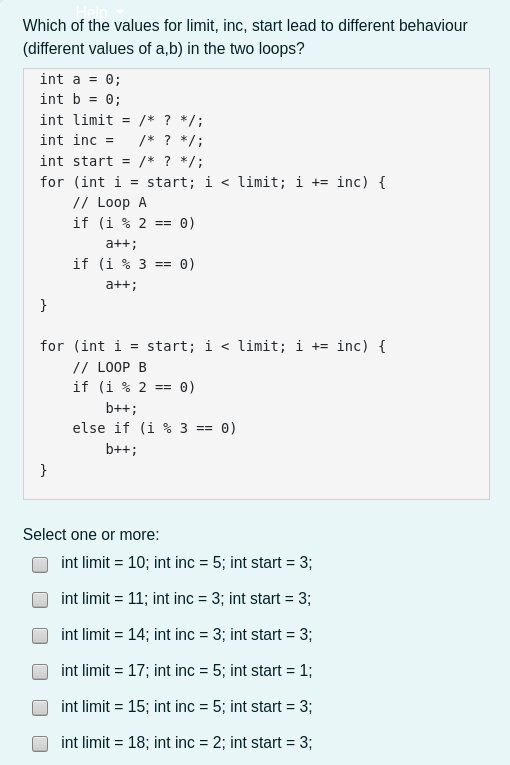}
  \end{minipage}
  
(a) \hspace{0.45\linewidth}(b)
  \caption{Multiple choice question asking to count loops (a) and to distinguish successive conditionals from an if-then-else (b).}
  \label{fig:tuwelImg}
\end{figure}

Control flow tracing exercises consist of small computer-program sources that can be used to test a student's ability to read and understand given code.
Two examples for a control flow tracing exercise in the Java programming language can be found in Figure~\ref{fig:tuwelImg}.

In general, a tracing question requires the student to find the values of a particular set of variables or what output the given program has.
Tracing may also be used indirectly: the counters in Figure~\ref{fig:tuwelImg}(a) correspond to the number of times a loop is executed, but the student
only needs to order the loops by their execution length.

\begin{figure}[h]
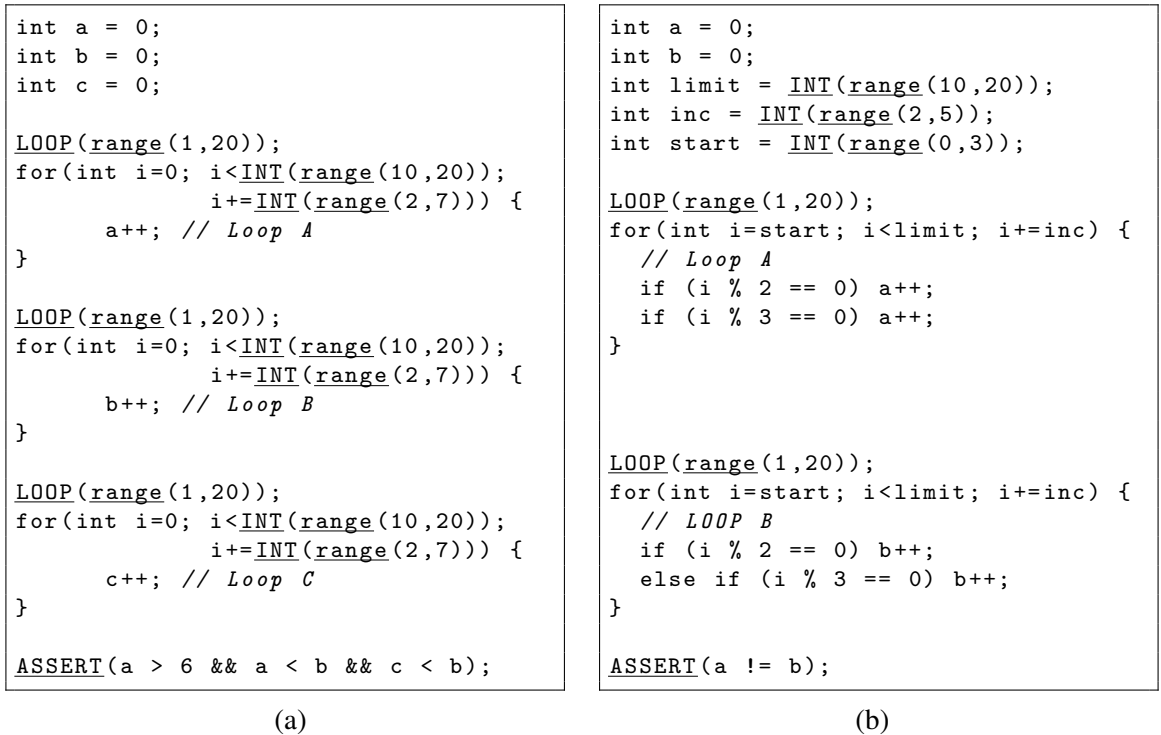

  \centering
  \begin{minipage}[t]{0.45\linewidth}
    \footnotesize
    \begin{lstlisting}[basicstyle=\ttfamily]
int a = 0;
int b = 0;
int c = 0;

LOOP(range(1,20));
for(int i=0; i<INT(range(10,20));
             i+=INT(range(2,7))) {
      a++; // Loop A
}

LOOP(range(1,20));
for(int i=0; i<INT(range(10,20));
             i+=INT(range(2,7))) {
      b++; // Loop B
}

LOOP(range(1,20));
for(int i=0; i<INT(range(10,20));
             i+=INT(range(2,7))) {
      c++; // Loop C
}

ASSERT(a > 6 && a < b && c < b);
\end{lstlisting}
  \end{minipage}
\hspace{5mm}
  \begin{minipage}[t]{0.45\linewidth}
    \footnotesize
    \begin{lstlisting}[basicstyle=\ttfamily]
int a = 0;
int b = 0;
int limit = INT(range(10,20));
int inc = INT(range(2,5));
int start = INT(range(0,3));

LOOP(range(1,20));
for(int i=start; i<limit; i+=inc) {
  // Loop A
  if (i % 2 == 0) a++;
  if (i % 3 == 0) a++;
}



LOOP(range(1,20));
for(int i=start; i<limit; i+=inc) {
  // LOOP B
  if (i % 2 == 0) b++;
  else if (i % 3 == 0) b++;
}

ASSERT(a != b);
\end{lstlisting}
\vfill
\end{minipage}\\
(a) \hspace{0.45\linewidth}(b)

  \caption{Tracing exercises focusing on higher-level analysis of a program}
  \label{fig:complicatedexercises}
\end{figure}

In a similar spirit, the program in Figure~\ref{fig:complicatedexercises}(b) tests the understanding of successive and nested branching. In cases where the loop variable
is divisible by 6 the two loops behave differently. The student receives a list of initialization blocks and has to mark those where the divergence
results in a different computation between the programs.

To generate variations of these exercises, we annotate the skeleton of a Java program with constraints on the values of variables and
assertions over some values. Often the assertions express a post-condition of the whole program, but they may be inserted in between statements as well.
Figure~\ref{fig:complicatedexercises} shows the skeletons used to generate the questions above.

\begin{figure}[h]
\begin{tabularx}{\textwidth}{|X|X|}
\hline
Given & Possible Question\\
\hline
A complete code fragment & ``What is the result/output of the given code?''\\
\hline
& ``Does the result/output change if we replace line \ldots by \ldots?''\\
\hline
A set of complete code fragments & ``Which of the given code fragments produces the result/output \ldots?''\\
\hline
& ``Which of the given code fragments produce the same (concrete) results/outputs \ldots?''\\
\hline
A code fragment containing a hole (\texttt{??}) & ``Which value must be inserted instead of \texttt{??} such that the loop iterates \ldots times/the result is \ldots?''\\
\hline
\end{tabularx}
\caption{Possible questions}
\label{fig:questions}
\end{figure}

\section{Question Types}
Classically, a tracing question has a text field where students enter the computed result. The validation of the answer is problematic though. The input needs to be validated due to ambiguous solutions (e.g.~\texttt{0x0F} and \texttt{15} may be used interchangeably).
While this problem might still be avoided by adjusting the exercise, the way e-learning platforms like Moodle validate free-form exercises by comparing the answer to a given list of possible solutions is quite limiting.~\footnote{~This is mostly a security requirement of the platform because general input validation would need the execution of arbitrary programs. The Moodle plugin Coderunner (\url{https://moodle.org/plugins/qtype\_coderunner} implements a secure environment for such tasks but it is not part of the default distribution. Often the installation of such a plugin is out of the control of the teaching such that we can not expect it to be present.}
For example, a task asking to fill in the missing part of a program can not be expressed due to the unbounded number of possible code-snippets that could be entered. In general, questions that have an infinite set of possible answers (e.g. where any string of an appropriate length is a correct solution) lead to similar problems. For this reason, we focus on questions where all solutions can be pre-computed. Often a subset of these solutions is presented in a multiple-choice form but under these limitations, free-form answers are also possible.

Another issue is that code can be easily copy pasted into an IDE and run instead of manually computing the result. Turning the code fragment into an image makes this approach harder but not impossible: a student might use OCR software to recover the text or they might simply type it into an editor.

Alternatively, we could adapt the question by asking the student to analyze multiple code fragments. Instead of computing the result they select those fragments that exhibit a particular behaviour. A possible question could propose multiple implementations for binary search and ask the students to select the correct ones.

Since correctness proofs are undecidable in general, a BMC based approach is likely to fail for such questions. But the example in Figure~\ref{fig:complicatedexercises}(b) presents a compromise: we leave holes for literals marked by \texttt{??} in the program and ask for values that lead to different behaviour of the two loops. The answer candidates are generated in two runs, the first creates the correct solutions and the second the wrong solutions. 

The table in figure~\ref{fig:questions} gives an overview of the question types we consider.

\section{Specifying a Program Skeleton}
The exercises are intended for beginners knowing only a restricted subset of the Java programming language. Therefore we restrict ourselves to what Lister calls a ``Pascal-like'' language with conditionals (if-else, ternary operator), loops (while, do-while, for), restricted jump statements (return, break, continue), and function calls (including recursion) as control flow operators. Data types are usually booleans, integers and arrays of integers. Further types the tool can deal with are strings, one dimensional string arrays and two dimensional integer arrays. However, reasoning about these kind of datatypes is often hard for solvers and the computation time consequently quite high for larger code-snippets.

\begin{figure}[h]
  \centering
  \begin{minipage}{\textwidth}
  \footnotesize
\begin{lstlisting}[basicstyle=\ttfamily,captionpos=b,label=lst:templateA]
@MAIN
static int[] start() {
  int[] input = INTARRAY(list(12), range(-25, 25));
  ASSERT(__distinct(input, 12));
  return mystery(input, 0, input.length - 1);
}

@REC(5)
static int[] mystery(int[] data, int from, int to) {
  if (from == to) return new int[] { data[from], data[from] };

  int[] left = mystery(data, from, (from + to) / 2);
  int[] right = mystery(data, (from + to) / 2 + 1, to);

  int r[] = new int[2];
  r[0] = left[0] < right[0] ? left[0] : right[0];
  r[1] = left[1] < right[1] ? right[1] : left[1];
  ASSERT(Math.abs(r[0] - r[1]) > 10);
  return r;
}
\end{lstlisting}
\end{minipage}
\caption{A recursion that computes minimum and maximum of a given input.}
\label{fig:recursiveexample}
\end{figure}

A program skeleton is a syntactically correct Java program with additional functions and annotations with special meaning. To differentiate the two we underline the additional functions in our examples. Processing the template in \tatsu{} results in one or multiple instances that fulfill the expressed constraints. An instance is a plain Java program without any annotations left.

The most important functions are the placeholders that represent concrete constant values in the generated instances. The placeholder itself represents the type (\texttt{INT}, \texttt{BOOLEAN}, \texttt{INTARRAY}, etc.) and describes the possible values via an argument that contains a constraint. The function \texttt{list} selects one of its arguments as possible value which is particularly suited for non-contiguous intervals. The function \texttt{range} stands for an integer value in the range of its arguments. For example, the loop bounds and increments in Figure~\ref{fig:complicatedexercises}(a) are specified as integers in the range of $[10;20]$ and $[2;7]$ respectively. In Figure~\ref{fig:recursiveexample} the placeholder \texttt{INTARRAY} represents an \texttt{int[]} of size $12$ containing values in the range $[-25; 25]$.

Furthermore, we can also restrict the values for the placeholders indirectly by adding assertions with \texttt{ASSERT}.
In Figure~\ref{fig:recursiveexample} we have two such assertions: the one in \texttt{mystery} guarantees that the difference of maximum and the minimum in each recursive step is at least $10$. The assertion in \texttt{start} ensures that the generated array does not contain duplicate elements by using the helper function \texttt{\_\_distinct}.

Other important constructs are the \texttt{LOOP} and \texttt{@REC} annotations which limit the unwinding of loops and recursions. This is motivated by our goal of translating into a decidable (quantifier-free) logic theory suited to bounded model checking\footnote{~The generated problem might still be undecidable due to the assertions. Examples for this are constraints on strings or when explicitly using quantifiers but this is under control of the user.}. The
\texttt{LOOP} annotation describes the number of iterations a loop may be repeated. This is often a range with lower bound zero and a reasonable upper bound. However, to guarantee a minimum complexity for a student to trace, a different lower bound on the interval or even a fixed list of possible repetitions may be specified. For example, the loops in Figure~\ref{fig:complicatedexercises} need to iterate at least once and up to twenty times.

The \texttt{@REC} annotation describes the maximum amount of recursive calls of a function for the current trace. In contrast to \texttt{LOOP} the \texttt{@REC} annotation is optional and defaults to non-recursive calls only. For example, the maximal recursion depth of \texttt{mystery} in Figure~\ref{fig:recursiveexample} is set to five.

These limits can be thought of as constraints on the desired models. Similar to overly strict assertions, \tatsu{} will simply report that it could not find a model in these cases.

Assertions that can not be represented by a single statement/assertions can be wrapped into an \texttt{ASSERTBLOCK}. This statement indicates that the following block should be considered as a ``large assertion'' and not as code that will be handed out to the students.

The skeleton contains two different kinds of code: code that is present in the generated snippet and code that is only considered during generation. An example can be seen in Figure~\ref{fig:preprocessing}. The \texttt{ASSERTBLOCK} there constraints the array \texttt{arr} such that all values but one are odd. This block is not suitable to be used as hole in a question but it improves the examples generated.

\begin{figure}[h]
\centering
\begin{minipage}{\textwidth}
\footnotesize
\begin{lstlisting}[basicstyle=\ttfamily,captionpos=b,label=lst:templateB]
int[] arr = INTARRAY(list(5), range(1, 100));
int idx = INT(range(0, 4));
ASSERTBLOCK();
{
	ASSERT((arr[idx] % 2) == 0);
	ASSERT(__distinct(arr, 5));
	LOOP(list(5));
	for (int i = 0; i < arr.length; i++) {
		ASSERT(__impl(i != idx, (arr[i] % 2) == 1));
	}
}
arr[idx] /= 2;
arr[idx] *= 2;
\end{lstlisting}
\end{minipage}
\begin{minipage}{\textwidth}
\footnotesize
\begin{lstlisting}[basicstyle=\ttfamily,captionpos=b,label=lst:templateC]
int[] arr = new int[] { 23, 8, 43, 67, 59 };
int idx = 1;
arr[idx] /= 2;
arr[idx] *= 2;
\end{lstlisting}
\end{minipage}

\caption{Large assertions}
\label{fig:preprocessing}
\end{figure}

In a limited number of cases we can also use invariant-assertions to generate instances that can take a large number of loop iterations into account. Consider the skeleton shown in Figure~\ref{fig:invariant}. Again, the actual value for \texttt{INT(range(1, 100))} is replaced by \texttt{??} and the student is asked to find a value such that the last element in the array has the value given in \texttt{res}. As the number of iterations is large students cannot simply trace the flow directly. They have to find out that in the $i^{th}$ iteration of the loop the value in the current array element is $inc \cdot i$ and that after \texttt{arr.length} iterations the value at the last array position is $inc \cdot (arr.length - 1)$. Internally, the invariant replaces the loop unwinding but adds the assertion that the invariant follows from the loop body. This approach is helpful in cases with very high loop counts that create enormous translations which would not terminate in an acceptable amount of time. In general the invariant may be an under-approximation of the program state after the loop. To avoid spurious models the user has to be very careful to specify a sufficiently strong invariant. In this case the safety net of running the generated program in Java described at the end of Section~\ref{sec:implementation} becomes obligatory.

\begin{figure}[h]
\centering
  \begin{minipage}{\textwidth}
  \footnotesize
\begin{lstlisting}[basicstyle=\ttfamily,captionpos=b,label=lst:templateD]
int[] arr = new int[INT(range(6000, 10000))];
int inc = INT(range(1, 100));
int i = 1;
INVARIANT(i >= 1 && i <= arr.length && arr[i - 1] == (i - 1) * inc);
while (i < arr.length) {
	arr[i] = arr[i - 1] + inc;
	i++;
}
System.out.print(arr[arr.length - 1] == INT(range(60000, 100000)));
ASSERT(__out.equals("true"));
\end{lstlisting}
\end{minipage}
\begin{minipage}{\textwidth}
\footnotesize
\begin{lstlisting}[basicstyle=\ttfamily,captionpos=b,label=lst:templateF]
int[] arr = new int[9546];
int inc = 8;
int i = 1;
while (i < arr.length) {
  arr[i] = arr[i - 1] + inc;
  i++;
}

System.out.print(arr[arr.length - 1] == 76360);
\end{lstlisting}
\end{minipage}

\caption{Invariant assertions}
\label{fig:invariant}
\end{figure}

\section{Implementation}
\label{sec:implementation}

\tatsu{} shares some ideas with automated unit test generators like KLEE~\cite{DBLP:conf/osdi/CadarDE08} or Java PathFinder~\cite{pathFinderI,pathFinderII}. There, we want to find inputs for a given program that should trigger as many different program behaviours as possible whereas we create multiple programs that should behave quite similar.
The strongest influence are programs like CBMC~\cite{DBLP:journals/sttt/ArmandoMP09, CBMC} where bounded model checking is used to find bugs in programs. Where CBMC usually checks all possible program traces, we are only interested in generating some witness traces to fill the holes in the program skeleton.
Nevertheless, the approach is similar in the sense that it finitely unwinds the program into a variant of a single static assignment form (SSA form) and uses a solver to find a model of the logical representation. The model values
are used to fill the placeholders in the program skeleton. Finally, we run the instantiated skeleton to check if none of the assertions is violated in an actual Java environment.

Our encoding of Java structures into first-order terms is based on a simplified Java specification. A value of boolean type maps to a propositional variable, integer types map to bitvectors (Figure~\ref{fig:javatypemapping}). Java strings values map to \smtlib{} strings. This, on the one hand, gives us the advantage of arguing about strings of an arbitrary size (as this requires no loop unwinding) but, on the other hand, might increase complexity quite a lot.

\begin{figure}[h]
  \centering
\begin{tabular}[h]{|c|c|c|c|c|c|c|c|c|}
\hline
  Java type   & boolean & byte & short & int & long & char & String \\
\hline
  Placeholder & BOOLEAN & & & INT & & CHAR & STRING \\ 
\hline
  \smtlib{} & Bool & BitVec 8 & BitVec 16 & BitVec 32 & BitVec 64 & Unicode\protect\footnotemark & String \\
\hline
\end{tabular}
\caption{Mapping of Java types to \smtlib{} types}
\label{fig:javatypemapping}
\end{figure}

\begin{figure}[h]
  \begin{subfigure}[c]{\textwidth}
    \centering
    \includegraphics[width=0.75\textwidth]{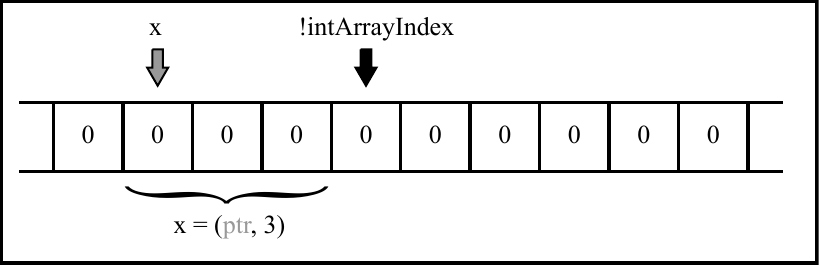}
    \caption{\texttt{!intArray} containing only some variable $x = \{ 0, 0, 0 \}$}
  \end{subfigure}
  \begin{subfigure}[c]{\textwidth}
    \centering
    \includegraphics[width=0.75\textwidth]{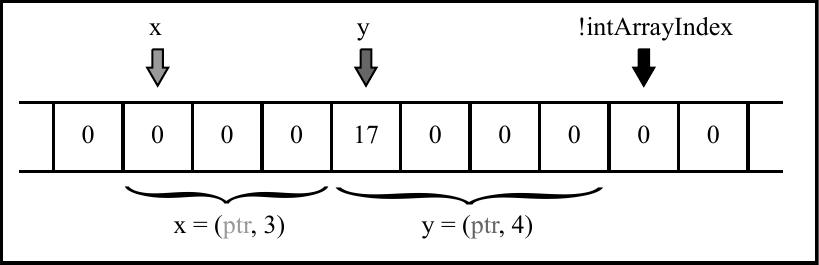}
    \caption{\texttt{!intArray} after executing \texttt{int[] y = new int[4]; y[0] = 17;}}
  \end{subfigure}
  \caption{Allocating memory}
  \label{fig:arrays}
\end{figure}

\footnotetext{~Z3 specific}Arrays are the only supported objects where we need differentiate between object and reference equality, as strings are readonly. Arrays are modeled as subsets of the heap that is  a single \smtlib{} array. A concrete Java array is represented as a pair of the index of the position in the heap and the arrays length. These tuples can be determined during unwinding since Java arrays neither overlap nor are they resizable. The heap is modeled explicitly as we have to allow two array variables to refer to the same memory address and writing to one of them has to change the value of the other one as well. From a logical perspective: For the SMT solver two arrays are considered equal iff their content is equal but in Java two arrays are equal iff the memory address they are referencing is equal.
Therefore, for every array type (\texttt{int[]}, \texttt{int[][]}, \texttt{String[]}) there exist global variables (in SSA form) representing the values on the heap for the corresponding type. For example, for \texttt{int[]} there exists a constant called \texttt{!intArray} that contains all the \texttt{int[]} values in the system as a subset. Furthermore, there exists another constant \texttt{!intArrayIndex} that marks the index in the global int array where arrays generated by \texttt{new int[y]} should be located. i.e., it represents the address of the generated array. This value will be incremented after the allocation accordingly to point to a free position again. Monotonically incrementing this index is a simple strategy that allows us to simulate allocating memory on the heap, however, forces us keeping track of variables that might have been already freed by Java. A concrete example how arrays are allocated is shown in Figure~\ref{fig:arrays}.

In case of multidimensional (``jagged'') arrays like \texttt{int[][]} we can make use of multiple global arrays. One refers to indices in the global \texttt{int[]} array and one contains the lengths of the corresponding arrays. The mapping of the 3 implemented array types is presented in Figure~\ref{fig:javaarraytypemapping}.

\begin{figure}[h]
  \centering
	\begin{tabular}[h]{|c|c|c|c|}
	\hline
	  Java type & int[\mbox{ }] & int[\mbox{ }][\mbox{ }] & String[\mbox{ }] \\
	\hline
		Placeholder & INTARRAY & INT2DARRAY & STRINGARRAY \\
	\hline
	  \multirow{2}{*}{\smtlib{}} & \multicolumn{3}{c|}{(BitVec 32, BitVec 32)} \\ 
	\cline{2-4}
	  & Array BitVec 32 & 3 $\cdot$ (Array BitVec 32) & Array String \\
	\hline
	\end{tabular}
\caption{Mapping of Java array-types to \smtlib{} types}
\label{fig:javaarraytypemapping}
\end{figure}

As in SMTLIB we use multi-sorted first-order logic with equality and if-then-else conditionals~\cite{BarFT-RR-17}. The theories used are bitvectors (BV), extensional arrays (AX), strings (S), datatypes (DT)~\cite{smtlib-theories, BarFT-RR-17}. The standard translation is quantifier-free but quantifiers can be added explicitly by the user if required (e.g., as part of invariants etc.). We also rely on custom extensions only present in Z3 (conversion between strings and bitvectors, constant arrays, etc.). A more formal definition of this section can be found in Appendix~\ref{sec:translation}.

Our tool converts the program's structure into SSA form. In order to do so, every possible change of the value of a variable requires us to introduce a new SMT constant. In order to do so, Java variables \texttt{x} are represented by a sequence of constants \texttt{x@[version]@[writecount]} where \texttt{[writecount]} is a number starting with $0$ that will be incremented every time there is an assignment to the variable \texttt{x}. As multiple different variables with the same name \texttt{x} might occur during translating the code (due to Java scoping, loop unwinding, function inlining, \ldots) the number \texttt{[version]} represents how many different variables with the same name have been processed before. Most SMT constants used internally include some characters like ``@'' or ``!'' to avoid collisions with user-defined variables. However, in some cases it is not possible as the internal variables needs to be reflected in the Java code. An example for such a name is the pseudo-variable \texttt{__out} representing the content printed to console so far. It can be used in assertions to constraint the output and therefore needs a Java-compatible name. The output command \texttt{System.out.print(s);} is then simply modeled by \texttt{__out += s;}.

Among the control constructs, the \texttt{if-then-else} conditional translates to two logical implications expression whereas \texttt{(do-)while}- and \texttt{for}-loops are removed during unwinding. The \texttt{break}, \texttt{continue}, and \texttt{return} statements are also handled directly during the SSA form generation by a program transformation.

In order to introduce only as few additional SSA-constants as possible we slightly modify the SSA approach. The idea of the SSA-form is that every variable assignment introduces a new constant~\cite{SSA}. In case of branching (due to a simple if-then-else or introduced by loop unwinding) the ordinary approach would use disjoint constants in both branches. However, the approach used here makes only sure that the solver does not have to assign different values to the same constant. Some SMT constant \texttt{x} might, therefore, occur in both the then- and the else-branch. Figure~\ref{fig:ssa} shows both approaches.

\begin{figure}[h]
\begin{tabular}{|c|c|c|}
\hline
Java code & Ordinary SSA & Modified SSA \\
\hline
\begin{lstlisting}[breaklines=true,frame=none]
int x = init;
if (cond) {
  x += v1;
  x -= v2;
} else {
  x *= v3;
}
int y = x + 1;
\end{lstlisting}&
\begin{lstlisting}[breaklines=true,frame=none]
x@0 = init;
x@1 = x@0 + v1;
x@2 = x@1 - v2;
x@3 = x@0 * v3;
x@4 = cond ? x@2 : x@3;
y@0 = x@4 + 1;
\end{lstlisting}&
\begin{lstlisting}[breaklines=true,frame=none]
x@0 = init;
if (cond) {
  x@1 = x@0 + v1;
  x@2 = x@1 - v2;
} else {
  x@1 = x@0 * v3;
  x@2 = x@1;
}
y@0 = x@2 + 1;
\end{lstlisting}\\
\hline
&\begin{lstlisting}[breaklines=true,frame=none]
(and
 (= x@0 init)
 (= x@1 (+ x@0 v1))
 (= x@2 (- x@1 v2))
 (= x@3 (* x@0 v3))
 (= x@4 (ite cond x@2 x@3))
 (= y@0 (+ x@4 1))
)
\end{lstlisting}&
\begin{lstlisting}[breaklines=true,frame=none]
(and
 (= x@0 init)
 (ite
  cond
  (and
   (= x@1 (+ x@0 v1))
   (= x@2 (- x@1 v2)) 
  )
  (and
   (= x@1 (* x@0 v3))
   (= x@2 x@1)
  )
 )
 (= y@0 (+ x@2 1))
)
\end{lstlisting}\\
\hline
\end{tabular}
\caption{SSA translation}
\label{fig:ssa}
\end{figure}

As we can see, our representation is slightly larger in general than the ordinary one (w.r.t. \smtlib{} code size). However, it has two advantages over its alternative. Firstly, it introduces less constants (especially if both branches contain multiple write accesses to the same variable). This allows us to directly read off the concrete data-flow from the generated model which is particular interesting in potential post-processing steps. Secondly, we preserve the original structure of the Java code which can be useful in case the generated problem encoding has to be modified afterwards.

Post-processing, maintainability, and readability play an important role in the underlying encoding for us. Transforming and simplifying the generated code appears to be a reasonable idea.
Our code optimizations lead from simple arithmetic pre-computations (for example, replacing $1 + 2$ by $3$) to keep track of possible values of variables via an interval abstraction. Doing these simplifications can be used to eliminate unreachable branches or to detect better loop-unwinding bounds. An example for the latter can be found in Figure~\ref{fig:unwinding} where the loop limit is signifcantly larger than required:
 The variable \texttt{count} might be initialized by some value in the range $[0, 10]$ and \texttt{i} has some value in the range $[3, 5]$. Iterating one time will result in \texttt{count} being contained in $[1, 11]$ and \texttt{i} between $[5, 7]$. We can update the intervals until the loop condition is false or the regular upper limit is reached. In our example, this happens after the seventh iteration where the interval $[8, 18]$ for \texttt{count} and the interval $[19, 21]$ for \texttt{i} do not intersect anymore. From this point on, unwinding the loop further does not make sense because at least at this point the loop condition is invalidated.

\begin{figure}[h]
  \centering
  \begin{minipage}{\textwidth}
  \footnotesize
\begin{lstlisting}[basicstyle=\ttfamily,captionpos=b,label=lst:templateE]
int i = INT(range(3, 5));
int count = INT(range(0, 10));
LOOP(range(0, 100000));
while (i < count) {
	 count++;
	 i += 2;
}
\end{lstlisting}
	\end{minipage}
	\caption{Detection of loop bounds by unwinding.}
	\label{fig:unwinding}
\end{figure}

In ordinary software there are mostly no (reasonable low) loop iteration bounds. However, when considering small exercises for students with bounded constants we can often completely unwind loops. None of these simplifications can be understood as an assistance for the solver. Most solvers have very sophisticated preprocessors and would probably do much more powerful optimizations on their own. We ran all our Unit-Tests with and without simplification. There was at most only a very slight performance improvement when optimization was turned on. But this was not the point of doing these kind of simplifications. The main reason for applying them is to generate a short, succinct, and as readable as possible representation of the input program. Apart from that, is mostly easier and more convenient for persons writing program skeletons to let the program detect the loop bounds itself instead of estimating them by their own unless they explicitly want to control the number of iterations. The creators of the program skeletons should only have to get in touch with the underlying technology/logic to the extent that is absolutely necessary. The tool should not require its users to understand what is happening internally.

We then use the SMT solver Z3 to find models of the generated formula. Each model value is then translated back to a Java expressions. The conversion of booleans, characters, integer, and string types back to Java is (almost) direct whereas arrays are constructed as sub-arrays of the heap.
These expressions are inserted for the placeholders of the skeleton to create the Java instance. Multiple instances can be generated by excluding the placeholder values found so far by adding appropriate blocking clauses.

As a safety net we double check the programs produced by adding the skeleton's constraints as Java assertions and running the generated instance. This provides a second source for both the output and return values of the computation.
Doing this check also gives us a method to guarantee the correctness of our found instance and to detect bugs in the generator may it be due to translation mistakes or a different understanding of the Java specification. (The behaviour of the Java standard library may change between versions\footnote{~For example the expression \texttt{"" == "abc".substring(0,0)} evaluates to false in OpenJDK prior to version 15, but it evaluates to true in current versions. This is due to an optimization of the substring function made in the bug report JDK-8240094 (\url{https://bugs.java.com/bugdatabase/view_bug.do?bug_id=JDK-8240094}).}.) 

\section{Outlook and Limits of the Method}

We are currently experimenting with further optimizations that can reduce generation time. In particular code doing manipulations on (nearly) all elements of an array (e.g., sorting) is very hard to deal with w.r.t. performance. We consider floating-point types out of reach both due to the difficulty of their logical encoding~\cite{DBLP:conf/arith/BrainTRW15} and the limited contribution to understanding precise float computation by tracing. Furthermore, the generator currently misses object oriented aspects like objects and class variables.

A theoretical problem that might occur is that we cannot guarantee that all possible models have the same probability of getting chosen. We do not really have much control over what model is found by the solver and it might happen that the solver yields very similar models every time. For example, the expression \texttt{INT(range(0, 1000))} might become $0$ in the first model, then $1$, then $2$, and so on. In practice, we have not encountered such strong dependencies between the generated models. However, this depends on the internals of the used solver.

\bibliographystyle{eptcs}
\bibliography{referencesPaper}
\newpage
\appendix
\section{Java Translation to FOL}
\label{sec:translation}
In the following we will describe a formal translation of the input program into a formula. The actual implementation (SSA, constant naming, etc.) of the translation is outlined in the main part of this paper. We restrict the formalisation to a smaller subset of the supported Java fragment due to space limitations.
A translation of Java variables to a logical constants needs to take two issues into account: assignments require the introduction of a fresh successor and the same name might refer to a different memory location in different scopes. A context collects the information necessary to perform this mapping.

\begin{definition}
  A \textit{context} is a tuple $\tuple{ N, V, E }$ where $N$ is a set of logical constants, $V$ is a sequence of mappings from Java variables to logical constants, $E$ is a set of logical formulas.
\end{definition}

The set $N$ tracks the logical constants introduced. A fresh constant is then just one that has not been introduced yet. The sequence of mappings $V = [v_1,\ldots,v_n]$ represents
the nesting of scopes where the innermost nesting has the highest index. A lookup successively widens the scope until it finds a mapping for the variable in question. A variable declaration updates the
innermost context.  Entering a new scope adds a fresh mapping to the sequence whereas leaving the scope removes the the last mapping from $V$. The set $E$ will track some side conditions such the handling of division by zero during the translation of expressions. We will also use the notion $\domain(f)$ for the domain of the function $f$ here.

This is formally expressed as follows:

\begin{definition}
  \noindent
  Let $C = \tuple{N,V,E,P}$ be a context with $V=[v_1,\ldots v_n]$. Then the function $\lookup{C}{jv}$ maps a a context and a Java variable to a logical constant and is recursively defined as:\\
  \[
    \lookup{C}{jv} = \left\{
        \begin{array}{ll}
          v_n(jv) & \text{if }jv\in \domain(v_n) \\
          \lookup{\tuple{N, [v_1,\ldots,v_{n-1}],E,P}}{jv} & \text{if }jv\not\in \domain(v_n) \text{ and } n \geq 1\\
          \text{undefined} & \text{otherwise}\\
        \end{array}
      \right.
    \]

    \noindent
    Furthermore, let $x \not \in N$ be a logical constant, $N' = N \cup \set{x}$,
    $v_n'(j) = \left\{\begin{array}{ll}
                        x & \text{if } x = jv\\ 
                        v_n(j) & \text{otherwise}\\
                      \end{array}\right. $
                    be a variable mapping extended by a Java variable $jv$, and let $v_{n+1}$ be a function with an empty domain.\\

     \noindent
    Then we define the following functions:\\
    \begin{tabular}{l@{$\,=\,$}ll}
      $\update{C}{jv}$ & $\tuple{N',[v_1,\ldots,v_n'], E, P}$ & maps a context and a Java variable to a new context\\
    $\enter{C}$ & $\tuple{N,[v_1,\ldots,v_n, v_{n+1}], E, P}$ & maps a context to a new context\\
    $\leave{C}$ & $\tuple{N,[v_1,\ldots,v_{n-1}], E, P}$ & maps a context to a new context\\
    $\addconstraint{C, F}$ & $\tuple{N,V, E \cup F,P}  $ & maps a context and a formula to a new context\\
    \end{tabular}
    
    ~
    
    We also extend $\update{C}{JVs}$ to sets of Java variables $JV$ by successively applying $\update{C}{jv}$ for each $jv \in JVs$.
    \noindent
    All three functions are undefined for an empty sequence $V$.
\end{definition}

During the unwinding of a Java program we never rely on one of the functions being undefined. This is due to the following invariants of Java programs:
\begin{itemize}
\item Program execution always happens within a scope
\item Every opened scope is closed again
\item Variables have to be assigned before use  
\end{itemize}

\noindent
The following translation makes some further assumptions:
\begin{itemize}
\item \texttt{for} and \texttt{do-while} loops have been rewritten to \texttt{while} loops
\item Function bodies have been rewritten such that \texttt{return} only occurs as the last statement
\item Loop bodies have been rewritten such that \texttt{continue} and \texttt{break} do not occur anywhere
\item The variables  \texttt{__out} and \texttt{\return{}} are never declared
\item Recursion, strings and arrays have been left out for simplicity
\item There are no uninitialised variables
\end{itemize}

We can now define the translation (including the unwinding of loops) of expressions ($T_E$) and of statements ($T_S$).
Both functions return a tuple with a context and a logical expression. In the case of $T_S$ the expressions are always of type $Bool$ such that they can be used in logical formulas.

\begin{definition}
  We define the functions $\unwindE{C}{expr}$ and $\unwindE{C}{stmt}$ by mutual recursion based on the structure of $expr$ and $stmt$.\\\noindent
  For $T_E$ We distinguish on the shape of $expr$:
  \begin{itemize}
  \item $expr$ is a literal of types $char$, $byte$, $short$, $int$, $long$: $\unwindE{C}{expr} =  \tuple{C, num}$
    where $num$ is the numeral corresponding to expr. The type of $num$ is determined by the table in Figure~\ref{fig:javatypemapping}.
  \item $expr$ is Java variable $jv$: $\unwindE{C}{expr} =  \tuple{C, \lookup{C}{jv}}$
    
  \item $expr$ is a \tatsu{} placeholder $INT(range(lower,upper))$: let
    \begin{itemize}
    \item [] $C = \tuple{N,V,E}$
    \item [] $c \not \in N$ is a fresh logical constant. The type of $c$ is determined by the placeholder ($INT$, etc.) and the table in Figure~\ref{fig:javatypemapping}.
    \item [] $l$ is the numeral corresponding to $lower$, $u$ is the numeral corresponding to $upper$
    \item [] $C' = \tuple{N \cup \set{c}, V, E \cup \set{l \leq c, c \leq u }}$      
    \end{itemize}
    then $\unwindE{C}{expr} = \tuple{C', c}$.
    
  \item $expr$ is one of the arithmetic expressions $s \circ t$ with $\circ \in \set{+,-,*,/,<,>,<=,>=,==,!=}$: let
    \begin{itemize}
    \item[] $\tuple{C_1, ls}~=~\unwindE{C}{s}$,
    \item[] $\tuple{C_2, lt}~=~\unwindE{C_1}{t}$,
    \item[] $F = \left\{
        \begin{array}{ll}
          lt \lnequiv 0 & \text{if }\circ = / \\
          \top & \text{if } \circ \in \set{+,-,*,<,>,<=,>=,==,!=} \\
        \end{array}
      \right.$
    \end{itemize}
    then $\unwindE{C}{expr} = \tuple{\addconstraint{C_2}{\set{F}}, ls \oplus lt}$ where $\oplus$ is the  logical arithmetic operator
    corresponding to $\circ$. Note that the comparison operators are expressions of type $Bool$ i.e. they are formulas that can be used as constraints.
  \item $expr$ is a pre-increment operator ++a / -{}-a: let
    \begin{itemize}
    \item[] $C_1 = \update{C}{a}$
    \item[] $\lookup{C_1}{a} \lequiv \lookup{C}{a} \circ 1$  with $\circ \in \set{+,-}$
    \end{itemize}
    then $\unwindE{C}{expr} = \unwindE{\addconstraint{C_1}{E}}{a}$.
    
  \item $expr$ is a post-increment operator a++ / a-{}-:
    \begin{itemize}
    \item[] $\tuple{C_1, t} = \unwindE{C}{a}$
    \item[] $C_2 = \update{C_1}{a}$
    \item[] $E = \lookup{C_2}{a} \lequiv \lookup{C_1}{a} \circ 1$  with $\circ \in \set{+,-}$
    \end{itemize}
    then $\unwindE{C}{expr} = \tuple{\addconstraint{C_2}{E}, t}$.

  \item $expr$ is a user defined function call $f(arg_1,\ldots,arg_n)$ with body $B$: let
    \begin{itemize}
    \item[] $\tuple{C_1, t_1}~=~\unwindE{C}{arg_1}$,
    \item[] $\tuple{C_2, t_2}~=~\unwindE{C_1}{arg_2}$,
    \item[] \ldots
    \item[] $ \tuple{C_n, t_n} = \unwindE{C_{n-1}}{arg_n},$
    \item[] $C' = \update{\update{\ldots \update{\enter{C}}{arg_1}}{\ldots arg_n}}{\return}$
    \item[] $\tuple{C'', F} = \unwindS{C'}{B}$
    \end{itemize}
    then $\unwindE{C}{expr} = \tuple{\leave{\addconstraint{C'',\set{F}}}, \lookup{C'}{\return}}$.
  \end{itemize}

  For $T_S$ we distinguish on the shape of $stmt$:
  \begin{itemize}
  \item $stmt$ is the empty statement: then $\unwindS{C}{stmt} = \tuple{C, \top}$
  \item $stmt$ is a \tatsu{} assertion $\ASSERT(cond)$:
    then $\unwindS{C}{stmt} = \unwindE{C}{cond}$
  \item $stmt$ is a block $\{ stmt \}$: let
    \begin{itemize}
    \item [] $\tuple{C', F} = \unwindS{\enter{C}}{stmt}$
    \end{itemize}
    then $\unwindS{C}{stmt} = \tuple{\leave{C'}, F}$
  \item $stmt$ is a return statement $\returnstmt{expr}$: let
    \begin{itemize}
    \item [] $\tuple{C_1, t} = \unwindE{C}{expr}$
    \item [] $C_2 = \update{C_1}{\return{}}$
    \end{itemize}
    then $\unwindS{C}{stmt} = \addconstraint{C_1}{\lookup{C_2}{\return{}} \lequiv{} t}$
  \item $stmt$ is a sequence of statements $stmt_1; stmt_2$: let
    \begin{itemize}
    \item[] $\tuple{C_1, F} = \unwindS{C}{stmt_1} $
    \item[] $\tuple{C_2, G} = \unwindS{C_1}{stmt_2} $
    \end{itemize}
    then $\unwindS{C}{stmt} = \tuple{C_2, F \land G}$
  \item $stmt$ is an assignment $jv = expr$: let
    \begin{itemize}
    \item[] $\tuple{C_1, t} = \unwindE{C}{expr} $
    \item[] $C_2 = \update{C_1}{jv} $
    \end{itemize}
    then $\unwindS{C}{stmt} = \tuple{C_2, \lookup{C_2}{jv} \lequiv t}$
  \item $stmt$ is a variable declaration $type~jv = expr$ with $type \in \set{char, byte, short, int, long}$ : let
    \begin{itemize}
    \item [] $\tuple{C_1, t} = \unwindE{C}{expr}$
    \item [] $C_2 = \update{C_1}{jv}$
    \end{itemize}
    then $\unwindS{C}{stmt} = \tuple{C_2, \lookup{C_2}{jv} \lequiv t }$.\\
    Note that the only difference to an assignment is that $v_n(jv)$ during $\lookup{C}{jv}$ would be undefined here.
  \item $stmt$ is a conditional $\ite{cond}{stmt_1}{stmt_2}$: let
    \begin{itemize}
    \item[] $\tuple{C_0, F} = \unwindS{C}{cond} $
    \item[] $\tuple{C_1, G_1} = \unwindS{C_0}{stmt_1} $ with $C_1 = \tuple{N_1, V_1, E_1}$
    \item[] $\tuple{C_2, G_2} = \unwindS{C_0}{stmt_2} $ with $C_2 = \tuple{N_2, V_2, E_2}$
    \item[] $C_{3} = \tuple{N_1 \cup N_2, V, \set{F \impl E | E \in E_1} \cup \set{\neg F \impl E | E \in E_2}) }$
    \item[] $Changed = \set{jv | \lookup{C}{jv} \neq \lookup{C_1}{jv} \lor \lookup{C}{jv} \neq \lookup{C_2}{jv}}$
    \item[] $C_{4} = \update{C_{3}}{Changed}$
    \item[]
      $H_1 = \bigwedge\limits_{jv \in Changed} (F \impl \lookup{C_4}{jv} \lequiv \lookup{C_1}{jv})$\\
      $H_2 =  \bigwedge\limits_{jv \in Changed} (\neg F \impl \lookup{C_4}{jv} \lequiv \lookup{C_2}{jv})$
    \end{itemize}
    then $\unwindS{C}{stmt} = \tuple{C_4, H_1 \land H_2}$
  \item $stmt$ is a conditional $\ithen{cond}{stmt_1}$:
    then $\unwindS{C}{stmt} = \unwindS{C}{\ite{cond}{stmt_1}{\{\,\}}}$
    
  \item $stmt$ is a loop $\tloopr{lower}{upper}\while{cond}{stmt'}$: we distinguish on the values of $lower$ and $upper$:
    \begin{itemize}
    \item $upper = 0$: then $\unwindS{C}{stmt} = \unwindS{C}{\ASSERT( !cond)}$
    \item $0 < lower < upper$: let
      \begin{itemize}
      \item [] $stmt'' = \ASSERT(cond); \tloopr{lower-1}{upper-1}\while{cond}{stmt'}$
      \end{itemize}
      then $\unwindS{C}{stmt} = \unwindS{C}{stmt''}$
    \item $0 = lower < upper$: let
      \begin{itemize}
      \item [] $stmt'' = \ithen{cond}{\{ \tloopr{0}{upper-1}\while{cond}{stmt'} \}}$
      \end{itemize}
      then $\unwindS{C}{stmt} = \unwindS{C}{stmt''}$
    \end{itemize}
  \item $stmt$ is a print statement $\texttt{System.out.print(expr)}$:let
    \begin{itemize}
    \item $\tuple{C', t} = \unwindE{\update{C}{\texttt{\_\_out}}}{expr} $      
    \item $C' = \tuple{N,V,E}$
    \end{itemize}
    then  $\unwindS{C}{stmt} = \tuple{N,V,E \cup \set{\lookup{C'}{\texttt{\_\_out}} \lequiv str.++(\lookup{C}{\texttt{\_\_out}},t)}}$
  \end{itemize}
\end{definition}

Given an initial context $C = \tuple{\set{out}, [{ \texttt{\_\_out} \mapsto out} ], \set{ out \lequiv ""}}$ we can translate the program by evaluating
$\tuple{C', F}{} = \unwindS{C}{ \texttt{main(new String[]\{\})}}$ with $C' = \tuple{N,V,E}$.\\
The specification is then $F \land E$.

\end{document}